# "Dynamic Ventilatory circuit divider": Safer way to multiply the number of available ventilators.

[1]Rashid Mazhar[*], [2]Muhammad E. H. Chowdhury, [1]Ahmed Faidh Ramzee, [1]Tasleem Raza, [1]Abdulaziz Ahmed Al-Hashemi.

[1]Pulmonology-Thoracic Surgery, Hamad General Hospital, Doha-3050, Qatar
[2]Department of Electrical Engineering, Qatar University, Doha-2713, Qatar

**\*** Correspondence: Dr. Rashid Mazhar; rashmazhar@hotmail.com, POBox 3050, Department of Pulmonology-thoracic surgery, Hamad Hospital, Doha, Qatar. Tel: +974-55546920

Dr. Rashid Mazhar. Sr. Consultant Pulmonology-Thoracic surgery, Hamad Hospital, Doha, Qatar.

Dr. Muhammad E. H. Chowdhury. A/P Electrical Engineering Dept. Qatar University.

(mchowdhury@qu.edu.qa)

Dr. Ahmed Faidh Ramzee, Resident surgery, Hamad Hospital, Doha, Qatar. (Faidhramzee@live.com)

Tasleem Raza, Sr. Consulatnt Pulmonolgy-Intensive care, Hamad Hospital, Doha, Qatar.

(Tmohd1@hamad.qa)

Abdulaziz Ahmed Al-Hashemi. Sr. Consulatnt , Pulmonary & sleep medicine, Hamad Hospital, Doha,

Qatar. (sleepmed@gmail.com)

**Conflict of Interest:** The authors declare no conflict of interest.

**Funding:** No formal internal or external funding was used during this work. No payment or monetary assistance was used for writing up this work.

**CRediT statement:**

**Rashid Mazhar:** Conceptualization, Methodology, Investigation, Writing- Original draft preparation, Supervision. **Muhammad E. H. Chowdhury:** Methodology, Investigation, Software, Data curation, Validation, Writing - Review & Editing. **Ahmed Faidh Ramzee:** Project administration, Resources, Validation. **Tasleem Raza:** Validation, Writing - Review & Editing. **Abdulaziz Ahmed Al-Hashemi:** Resources, Validation, Writing - Review & Editing.

# Abstract

**Background:**

Recent pandemic has brought a sudden surge in the requirement of mechanical ventilators all over the world. In this backdrop, there is wide interest in looking for ways to support multiple patients from a single ventilator. Various solutions, based upon simple mechanical division of the ventilator tubings are described. However, recently warnings have been issued by multiple international professional societies against these plumbing solutions as it can seriously harm the patients.

**Methods:**

We have bifurcated the inspiratory and expiratory conduits from a single ventilator with addition of one way valves, pressure and flow sensors along with volume and PEEP control. A purpose-built software and a low-cost microcontroller based control system integrates and displays the data in the familiar ventilatory graphic and numerical format onto a generic screen. The system is calibrated with simulated lungs with varying compliance.

In addition to the standard microbial, heat and moisture exchanger filters we design to add UV-C lights at 254-260 nanometre wavelength in the expiratory channel for its virucidal effect.

**Results:**

The "dynamic ventilatory divider" system is capable of providing and controlling individual flow, tidal volume (TV) and Positive End-expiratory pressure (PEEP) for individual patients. Furthermore, it would also display the ventilatory parameters of both the patients, on a single split screen, in a familiar format. FiO2 and rate are still controlled by the mother ventilator.

**Conclusions:**

The prototype system has a potential to provide safe ventilation to at least two individuals from a single ventilator, while maintaining the unique requirements of each patient.

**Introduction:**

With the ongoing pandemic of COVID-19, mechanical ventilators have become a rate-limiting therapeutic tool[1]. Manufacturing large numbers of ventilator machines is time consuming, besides the economic consideration in low-income countries. To fulfil this shortage, the popular media and some professionals have propagated an immediate solution by simple division of the ventilator tubings with suitable plumbing of the ventilator tubing with suitable connectors. Historically, this type of passive mechanical, divider solution is described in literature. These endeavours/reports fall in three categories: In-vitro experiments on lung simulators, In-vivo animal experiments and human study.

*A) In-vitro experiments on lung simulators*

Sommer et al.[2] and Neyman et al.[3] describe simple connectors dividing the ventilator circuit on plumbing fashion, without any individual provisions of pressure, volume sensing, control and display. These tests were carried with the presumption of equal ventilation to all four lung-simulators, presuming equal lung physiology. Branson[4] wrote a letter to the editor in criticism to the study reported by Neyman et al.[3]. Moreover, Branson et al.[5] have reported similar experiments to monitor individual parameters where the divided circuits were attached to dedicated pneumotachograph connected to an individual respiratory monitor.

The basic fact to be noted here that gas flow would follow the path of least resistance. The tidal volume for individual patients would be distributed according to the variability of compliance of lung rather than theoretical equal distribution. This type of apparent solution will cause severe lung injuries rather than supporting the patients as better lung (with higher compliance) will get more tidal volume and the stiffer lung (with lower compliance) will get less volume even though the requirement might be completely opposite.

Other critical respiratory problem and a cause of concern was the lack of control of positive end-expiratory pressure (PEEP). Hence, the apparently attractive concept of supporting multiple patients from a single ventilator by simple division/branching of the pipes is flawed and dangerous.

*B) In-vivo animal experiments*

Paladino et al.[6] carried out animal study using same simple division concept on healthy sheep for 12 hours. Again, there were no individual provisions of pressure, volume sensing, control and display. Furthermore, they were criticised[7] as being good only for supporting paralyzed, size-matched, and sedated individuals with normal lungs who are hemodynamically stable. The solution was deemed biologically irrelevant for mass respiratory failure, except possibly botulism (neurotoxicity with healthy lungs).

## *C) Human Study*

One published report [8] does describe the use of split ventilation in two volunteers (using a facemask interface rather than intubation). However, there was no provision for monitoring, controlling and display of individual ventilatory parameters.

Thus all the above, apparently one ventilator-multiple patients solutions remained out of practical use. With the dire current need of more and more ventilators, there is a resurgence in reporting of similar passive divider solutions.

However, this apparent simple divider solution is not recommended and recently warnings have been issued by individuals as well as multiple international professional societies against these as it can seriously harm the patients.[9-11]

The reasons against this "blind, uncontrolled division" of the ventilatory circuit include [9]:

- Inability to control individual volumes. With the branching circuits, the gas would follow the path of least resistance and the patient with stiffer lungs would take less and less volume while the patient with more compliant lungs would preferentially accumulate the incoming volume, harming both types of patients.
- Inability to manage individual PEEP.
- Inability to monitor individual patient's vital pulmonary mechanics.
- In case of sudden deterioration of a single patient (e.g., pneumothorax, kinked endotracheal tube), balance of ventilation would be distributed to the other patients.
- Risk of cross infection to all the patients attached to a single branching circuit.

It is suggested that to circumvent these shortcomings, with a volume-cycled mode, the deeply sedated patients having similar severity of lung injury, and preferably of the same body size are grouped together. The ventilator should be set up to pressure-cycled ventilation with a high PEEP and a low driving pressure (to achieve lung protection) and the ventilatory efficacy of each patient to be tracked using an end-tidal CO2 monitor placed in-line with their own endotracheal tube, allowing permissive hypercapnia. Furthermore, it is suggested that viral filters should be attached to prevent cross-contamination of pathogens between different patients [10].

Keeping these practical concerns in mind, we have devised the functional prototype of a "dynamic ventilatory circuit divider", whereby a single ventilator could be used to ventilate multiple patients, with *individual display and control* of the essential, required parameters of ventilation. Furthermore, it provides an anti-microbial provision, which is inherently anti-viral as well.

**Materials and methods:**

In the proposed system, mechanical, electronic and anti-microbial modifications were done to the ventilatory circuit. In principle, we have based our solution on the following functional aspects:

*I) Mechanical modifications*

*A) Flow direction in the branching circuits*: both the inspiratory and the expiratory circuits are provided with unidirectional flow valves to avoid cross-ventilation.

*B) Flow control*: Mechanical, gated valves were placed along the inspiratory conduits to graduate the tidal volume individually.

*C) PEEP control*: Mechanical, gated valves were placed along the expiratory conduits to graduate the PEEP in the target lungs. This is possible with the help of one-way valve placed in the inspiratory and expiratory limbs for individual circuit, which provides a closed path outside mother ventilator.

*II) Electronic modifications*

The mother ventilator along with its inspiratory and expiratory interfaces and display was interfaced to an add-on system, where an electronic control system and additional graphical display were used (Figure 1). This additional display can be computer, laptop or any other single-board computer with screen and universal serial bus (USB) interfaces. The electronic control system is a low-cost microcontroller based system, which is the brain of the dynamic bifurcated system and the graphical display was used as a user-interface only. In future, both of the system can be combined to have a control system along with graphical interface.

*A) Electronic sensors:* Two different sensors were used for each patient: Bi-directional flow sensor (SFM 3300-AW, Sensirion AG, Switzerland) and a Pressure sensor (MS4525-A5, TE connectivity). A bidirectional flow sensor placed distal (towards the patient) to the inspiratory control gate valve (Figure 1). A pressure sensor placed at the end of the expiratory tube along with the one-way valve (as shown in Figure 1).

*B) Graphical User-Interface (GUI):* The GUI shows the real-time flow, volume and pressure graphs and numerical values like a typical ventilator for individual patient. It immediately reflects any change happened to any of the divided system independently. A familiar digital display on a generic digital screen, which is divided to show both/all the patients ventilatory parameters on one screen, at the same time – akin to central monitoring display (Figure 2).

The innovation has the capability to a wireless display mode where multiple patients' management could be supervised by tale-medicine.

### *III) Anti-viral UV-C Light*

Furthermore, each individual circuit, coming out of the ventilator and the patients, is provided with a suitable standard bacterial filters as well as UV-C lights at 254-260 nanometre wavelength for its Virucidal effect. These tubing segments, carrying the UV light source are lined by internally reflective coating to prevent external scatter and plastic decay.

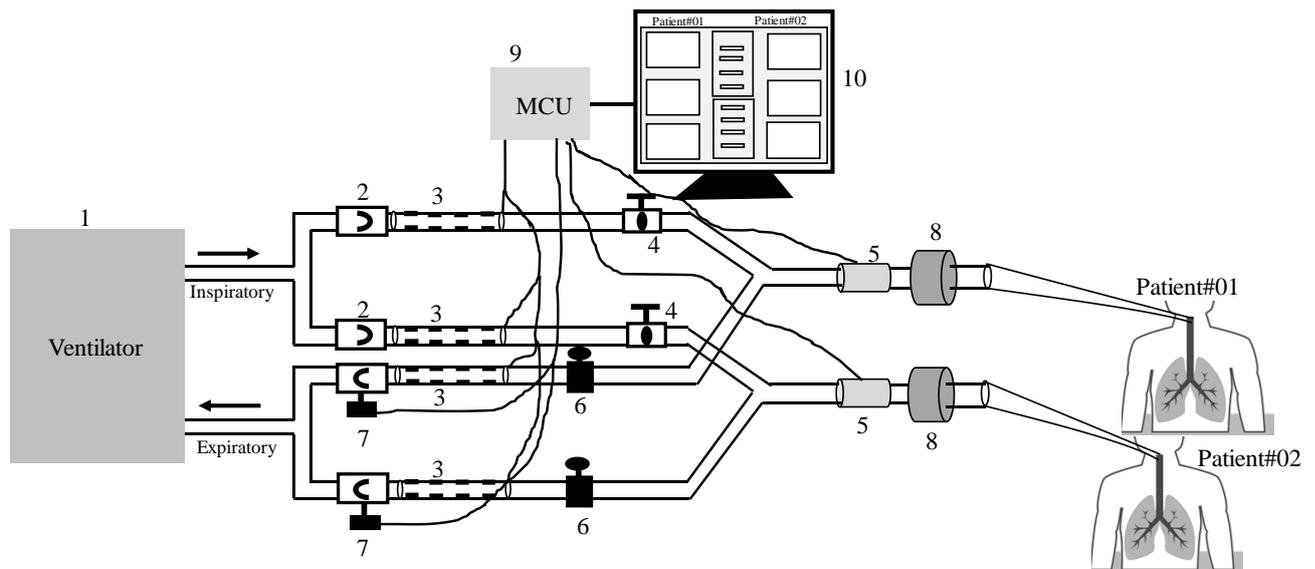

1. Mother Ventilator, 2. One-way valve, 3. Anti-viral UV light, 4. Volume Control, 5. Bi-directional flow sensor,
6. PEEP control, 7. Pressure sensor embedded in one-way valve, 8. HME filter 9. Microcontroller, 10. Graphical Interface (PC)

**Figure 1:** Simplified schematic diagram of a bifurcated, dynamic ventilatory circuit with its components.

*Functional considerations*

The control system of the prototype system gathers flow and pressure information from the sensors independently. The control unit is connected to the computer using USB interface to send data to the computer to display the individual patient parameters and graphs using the in-house built graphical user-interface using processing application. Both the flow and pressure sensors were calibrated using the mother ventilator while connected to two lung simulators of varying compliance (Figure-2). The calibration process is beyond the scope of this article, can be reported in a separate manuscript. All the conversion and calculation were done in the microcontroller and graphical user-interface (GUI) was used to provide an interactive but familiar interface to the respiratory therapists. Since the flow sensor is a bi-directional sensor, it can provide real-time inspiratory and expiratory flow directly in L/Min unit. This was directly used to show the real-time flow measurement in the GUI, however, the tidal, inspiratory and expiratory volume were calculated from the flow data in the microcontroller using the time of inspiration and expiration. On the other hand, the pressure sensor measures the pressure in the expiratory tube with respect to the air pressure and provides analogue voltage output according to the pressure. It was digitized using 10-bit analog-to-digital converter (ADC) of the microcontroller and calibrated to show the real-time pressure in cmH2O and PEEP pressure as numeric data.

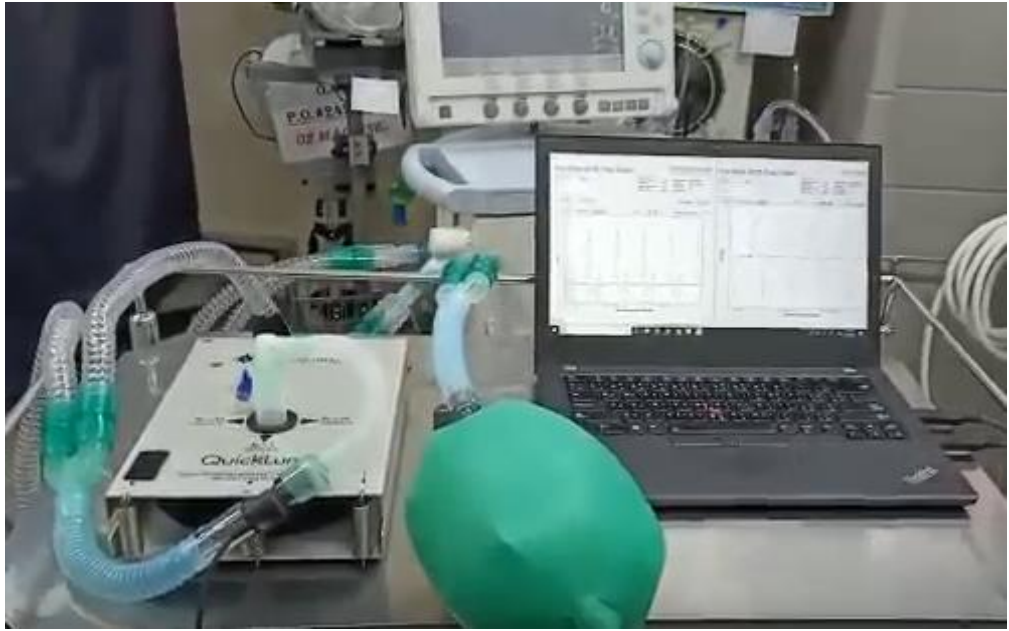

Figure 2: Initial testing and calibration of the prototype divider with two lung simulators of varying compliance.

## *Ventilatory considerations*

At the outset, the ventilator would be set on sedated patients, in volume control mode. Individual volume, flow and PEEP could be controlled in each branch. Rate and FiO2 would still be determined by the ventilator.

The divider circuit can modulate the Tidal volume of either patient. The ventilator would be set to deliver the total required cumulative volume for both the connected patients. The inflow to the patient requiring less TV would be brought to the required level by the flow control valve of that patient, leaving the other patient to receive the larger proportion of TV.

The ventilator would be set to deliver the lower of the two PEEP levels of the attached patient. If one patient's need higher PEEP, it would be increased to the required level by the PEEP control valve, leaving the other patient with the lower baseline PEEP.

The display of individual patients' real-time ventilatory data, graphic and numerical, is exactly in the format used by the standard ventilators (Figure-3) hence no new training needs to be given to the nurses, doctors and respiratory therapists.

**Results and Discussion:**

Our solution overcomes the "blind, uncontrolled division" approach suggested in the social media by a "dynamic ventilatory circuit divider" which provides each individual circuit branch, and hence each patient, with volume, pressures and flow rate sensing as well as controlling capability.

Two demo lungs' data/graphs were displayed on a single screen of a computer making it possible for one caregiver to manage both patients more ergonomically.

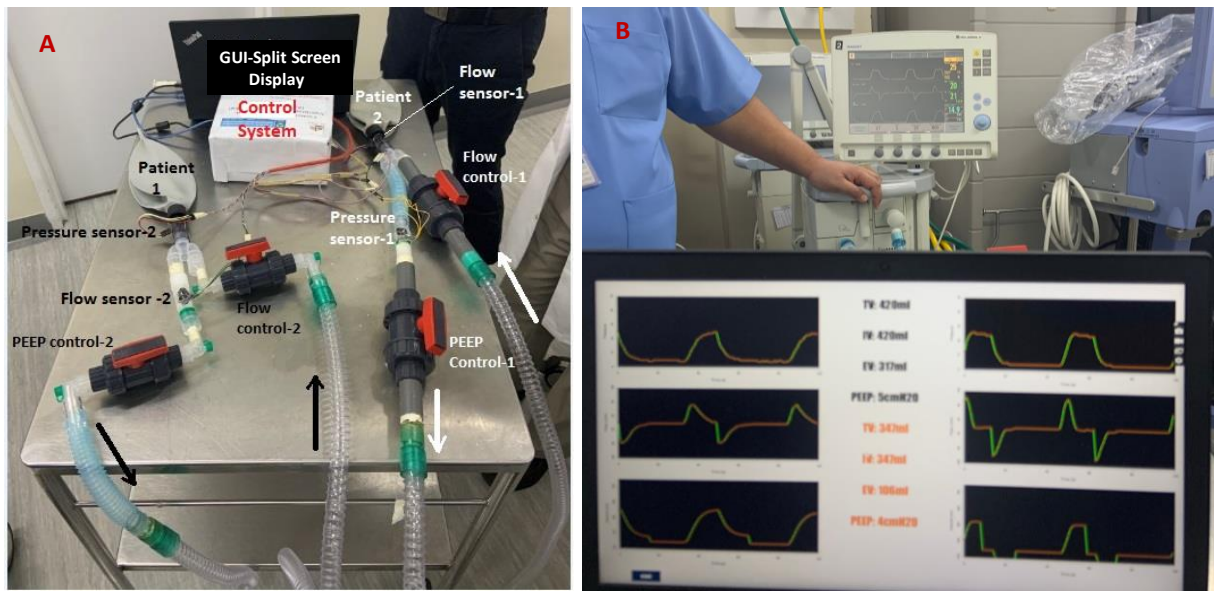

Figure 3: Snapshot of the complete "Dynamic Ventilation Circuit" with two-lung simulator (A), Snapshot of the working system with different Tidal volumes supplied to each demo patients (lung simulator) (B).

Different system components with the mechanical control, electronic sensing and control and user-interface is shown in Figure 3(A), while Figure 3(B) shows the dynamic bifurcated system is in operation. It can be clearly seen that individual patient's graphs and numerical data are shown side-by-side in the similar template of the conventional ventilator (Figure-3 B), which would make it easy to be used by the doctors, nurses and the respiratory therapists.

The proposed system has been evaluated for different clinically relevant scenarios to evaluate its functionality. Initially, mother ventilator was set to 800 mL tidal volume with peak pressure of 24 cmH2O and PEEP of 5 cmH2O. Although both the lungs were of similar compliance one lung was getting tidal volume of ~400 mL, while other is getting ~360 mL. This discrepancy is due to the imperfect circuitry of the prototype where the two limbs had slightly different lengths and ensuing dead space.

Figure 4 shows the graphic display of the individual control of TV of one patient whereby it could be tuned down to minimal, without affecting the tidal volume of the other patient.

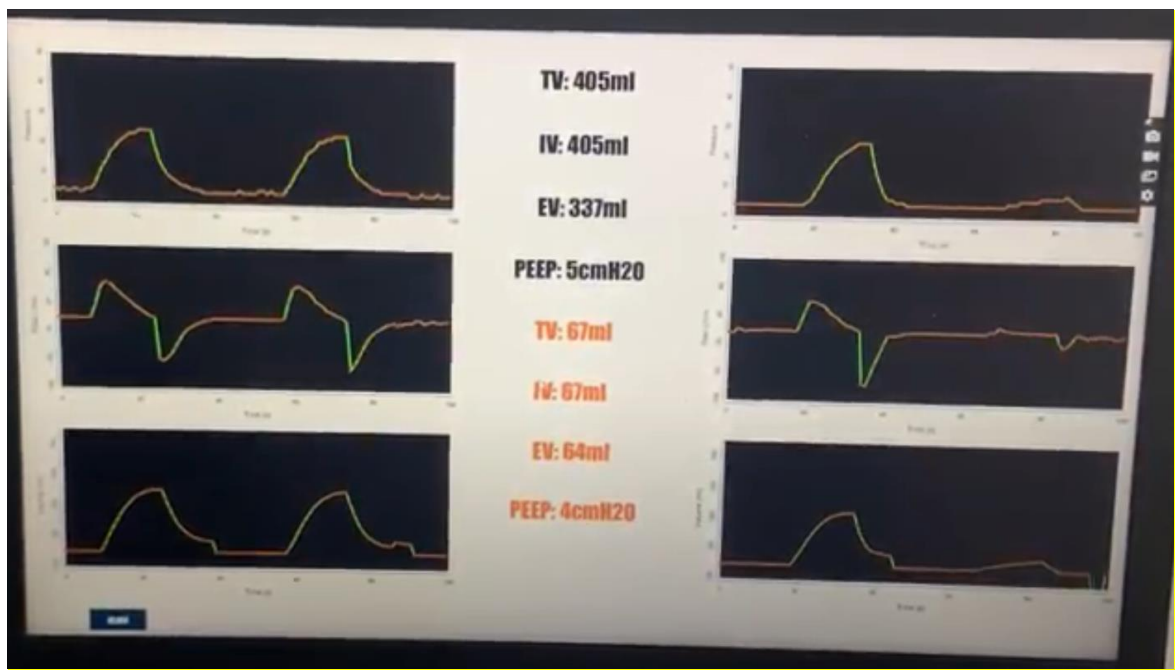

**Figure 4:** Demonstration of controlling and decreasing the Tidal volume of patient no. 2 (right) without affecting patient no. 1 (left).

Figure 5 demonstrates that the PEEP for one patient can be controlled individually without affecting the other patient. In this case, PEEP was increased to as high as 30 cmH2O while the PEEP for other patients was kept to 5cmH2O. It can be noticed that shape of pressure, flow and volume graphs for patient no. 2 (right) is unchanged while patient no. 1 (right) PEEP and volume has changed.

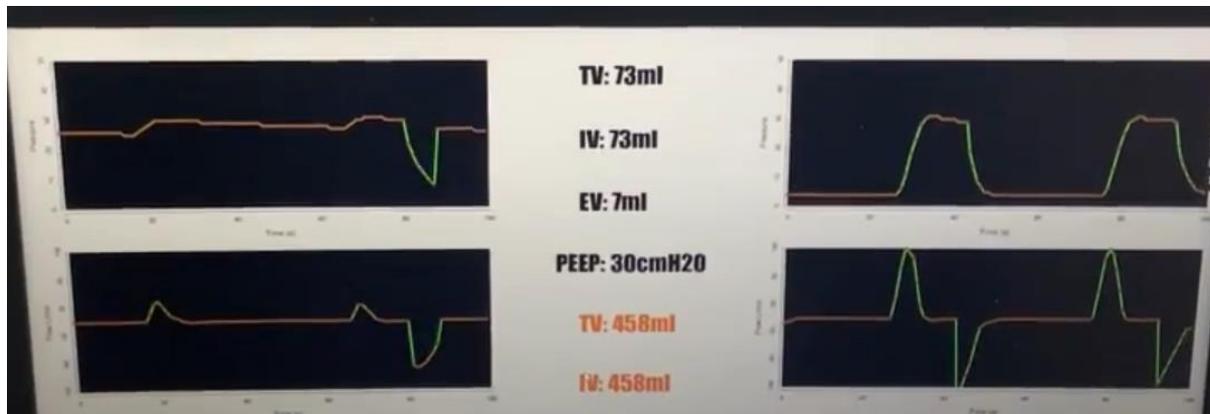

**Figure 5:** Demonstration of controlling PEEP to one (left) while all the parameters in the other patient (right) remains unchanged. Note: PEEP for patient one was modified in the beginning and then returned to normal during second cycle.

To measure the Ppeak and Pplateau pressures of one patient, the cumulative tidal volume from the ventilator would be dropped down to that particular patient's number, the flow gate of the other patient would be temporarily closed off and measurements would be taken from the ventilator in the usual manner. The sequence would be repeated for the other patient.

Our setup can be used for both pressure and volume controlled ventilation in controlled mode. Although the data in COVID-19 pulmonary involvement is still evolving, preliminary data in a significant proportion suggest a phenotypic pattern, which is different from classical acute respiratory distress syndrome (ARDS). These patients show severe hypoxemia and high dead space fraction, but lung compliance is relatively preserved. Therefore, patients generally do not require high PEEP, but high minute ventilation and oxygen supplementation. Non-invasive ventilation and high flow oxygen use in COVID-19 is discouraged, as they are high aerosol generating procedures and high infection risk for staff. Simple ventilatory modes may be adequate in these patients to provide oxygen support and therefore multiple-patient-single-ventilator may be a feasible strategy.

**Limitations & Conclusion:**

The provided solution is a method of dividing the ventilatory circuit with provision of monitoring, controlling and displaying the main ventilatory parameters of individual patients.

It is to be noted that this set-up is NOT being presented as a substitute for one ventilator-one patient solution, which remains the gold standard. The authors do see the only possible utility of this innovation in situations where there is a dire need to ventilate more patients than the available ventilators. Even under these circumstances, we DO NOT recommend attaching more than two patients to a single ventilator.

Minute ventilation requirements of patient may be drastically different, but this can be controlled with adjustment of tidal volume despite same respiratory rate for all patients. Patients will also require adequate sedation to avoid spontaneous breaths and ventilator patient asynchrony. During cardiac arrest situation, affected patient will require discontinuation from the circuit and main ventilator delivered volume will need readjustment for the remaining patient, in volume-controlled mode.

This is to re-emphasize that any and all such solutions are suboptimal as compared to a one patient-one ventilator setting. However, in the situation of gross mismatch of ventilator supply/demand, we believe that the presented solution is an economical, easy-to-adopt and relatively safer choice.

One suggestion of immediate utility could be application of this solution to stable, long-term ventilated patients in the society to free desperately needed ventilators for seriously ill patients.

**Acknowledgments:** The authors would like to thank the Electrical Engineering department, Qatar University and respiratory therapy unit, Hamad General Hospital for providing facilities and support to carry out the calibration of sensors, and conducting multiple testing of the system during this COVID-19 pandemic.

Individual thanks to Nelson D. Madriaga (Respiratory therapist, HGH), Mr. Iltaf Rabbani and Eng. Rakesh Gupta (HBK Eng-Med) and Anil Hapuarachchi (Q3DPrints) for their technical support as well as Eng. Umair Rashid for technical and writing support.

This manuscript, including related data, figures and tables has not been previously published and the manuscript is not under consideration elsewhere.

**Conflict of Interest:** The authors declare no conflict of interest.

**Funding:** No formal internal or external funding was used during this work. No payment or monetary assistance was used for writing up this work.

**CRediT statement:**

**Rashid Mazhar:** Conceptualization, Methodology, Investigation, Writing- Original draft preparation, Supervision. **Muhammad E. H. Chowdhury:** Methodology, Investigation, Software, Data curation, Validation, Writing - Review & Editing. **Ahmed Faidh Ramzee:** Project administration, Resources, Validation. **Tasleem Raza:** Validation, Writing - Review & Editing. **Abdulaziz Ahmed Al-Hashemi:** Resources, Validation, Writing - Review & Editing.

## Quick look

**Current Knowledge**

Several solutions claiming "one ventilator for multiple patients" have been publicised recently. Invariably these solutions depend upon simple division of ventilatory conduits by divider connections. Such an approach is not recommended as it would be dangerous for patients having different pulmonary compliance. The patients with more compliant lungs would get more and more volume at the expense of patients with stiffer lungs, harming both types.

**What This Paper Contributes To Our Knowledge.**

We describe a safer method of using one ventilator for at least two patients. A bifurcated ventilatory circuit is provided with an array of one way valves, pressure and flow sensors, along with flow controllers. A microcontroller and specific software processes the data to display the ventilatory parameters of each patient on a single generic screen. Thus, barring the FiO2 and the respiratory rate, the presented design allows volume, pressures and flow rate sensing as well as controlling capability for individual patients.